\documentclass[useAMS,usenatbib]{mn2e}
\usepackage{amssymb}
\usepackage{amsmath}
\usepackage{graphicx}
\usepackage{lscape}
\usepackage{color}
\usepackage{subfigure}

\def\h2{\rm{H_2}}
\def\fh2{f_{\rm{H_2}}}

\def\mh2{M_{\rm{H2}}}

\def\Sh2{\Sigma_{\h2}}

\title[Satellite galaxy quenching in SAMs]
{Resolution-independent modeling of environmental effects in semi-analytic models
of galaxy formation that include ram-pressure stripping of both hot and cold gas}

\author[Y. Luo]
{Yu Luo$^{1,2,4}$ \thanks{E-mail:luoyu,kangxi@pmo.ac.cn}, Xi Kang$^{1}$, Guinevere Kauffmann$^{2}$,
Jian Fu$^{3}$\\
$^1$Purple Mountain Observatory, the Partner Group of MPI f\"ur Astronomie, 2 West Beijing Road, Nanjing 210008, China\\
$^2$Max-Planck Institue f\"ur Astrophysik, 85741 Garching, Germany\\
$^3$Key Laboratory for Research in Galaxy and Cosmology, Shanghai Astronomical Observatory, Chinese Academy of Science,\\ 80 Nandan Road, Shanghai 200030, China\\
$^4$Graduate School, University of the Chinese Academy of Science, 19A, Yuquan Road, Beijing 100049, China\\
}

\begin{document}

\maketitle

\begin{abstract}

The quenching of star formation in satellite galaxies is observed over
a wide range  of dark matter halo masses and  galaxy environments. In the recent
Guo et  al (2011) and Fu  et al (2013) semi-analytic  + N-body models,
the gaseous  environment of  the satellite galaxy  is governed  by the
properties  of the  dark matter  subhalo  in which  it resides.   This
quantity depends of  the resolution of the  N-body simulation, leading
to  a  divergent   fraction  of  quenched  satellites   in  high-  and
low-resolution simulations.  Here, we incorporate an analytic model to
trace the subhaloes  below the resolution limit.   We demonstrate that
we then obtain  better converged results between  the Millennium I
and  II  simulations, especially for the satellites in the massive 
haloes ($\rm log M_{halo}=[14,15]$). We also include a new physical 
model for the ram-pressure stripping of  cold gas  in satellite  galaxies.
However, we find very clear discrepancies with observed trends in
quenched satellite galaxy  fractions as a function of  stellar mass at
fixed  halo  mass.  At  fixed  halo  mass,  the quenched  fraction  of
satellites  does  not  depend  on  stellar mass  in  the  models,  but
increases strongly  with mass  in the  data.  In addition to the 
over-prediction of low-mass passive satellites, the models also  
predict too few quenched  {\em central} galaxies with low stellar
masses,  so the  problems in  reproducing quenched  fractions are  not
purely of environmental origin.  Further improvements to the treatment
of the gas-physical processes  regulating the star formation histories
of galaxies are clearly necessary to resolve these problems.

\end{abstract}


\section{Introduction}

Many of  the observed properties  of galaxies, such as  their colors,
morphologies, star  formation rates  (SFR) and  gas-to-star fractions,
are  observed   to  have   strong  dependence  on   their  environment
(e.g.  Kauffmann  et  al.  2004;  Bamford  et  al.  2009;  Boselli  et
al. 2014). Galaxies in clusters or  groups tend to have redder colors,
bulge-dominated morphologies,  lower gas-to-star ratios and  less star
formation than  isolated galaxies of  similar mass (Butcher  \& Oemler
1978; Dressler  1980; Balogh et al.   2004; Baldry et al.  2006). This
dependence is  believed to  arise from the  interplay between  the gas
component  of  galaxies  and  their environment;  unlike  the  stellar
component, the gas  can be easily affected by the  ambient pressure in
galaxy groups or clusters.

When a galaxy moves through a cluster, the ram pressure (hereafter RP,
Gunn \&  Gott 1972) of  the intra-cluster  medium (ICM) acts  to strip
both the  hot gas reservoir  and the  cold interstellar gas,  and this
process plays an  important role in the star formation  history in the
galaxy. Many studies have concluded that this stripping process is the 
main cause for  the  increase  of  S0  galaxies  in  rich  clusters
(e.g. Biermann  \& Tinsley 1975;  Dressler 1980; Whitmore,  Gilmore \&
Jones 1993).  In the  last decade,  observations have  revealed direct
evidence for ram-pressure stripping of  gas in cluster galaxies in the
form of long gaseous tails  trailing behind these systems (e.g. Kenney
et  al. 2004;  Crowl et  al.2005;  Sakelliou et  al.2005; Machacek  et
al. 2006).  The same process is  also invoked as an explanation of the
depletion of the cold gas in  galaxies in clusters (Boselli \& Gavazzi
2006),  often  referred  to  as  ``HI  deficiency''  (e.g.  Haynes  \&
Giovanelli 1984; Solanes et al. 2001; Hughes \& Cortese 2009).

To understand the effects of RP  stripping on the galaxy gas component
in  detail,   numerical  hydrodynamical  simulations  are   the  ideal
tools. Abadi  et al. (1999) presented  the first study of  RP stripping
using an  idealized  SPH   simulation,  followed   by  more   realistic
hydrodynamical simulations  (e.g., Roediger \& Hensler  2005; Roediger
\& Br{\"u}ggen  2007; McCarthy et  al.  2008; Tonnesen \&  Bryan 2009,
Tecce  et  al. 2010).   These  studies  showed  that  RP can  strip  a
significant  amount of  hot gas  and cold  gas from  galaxies and  can
quickly reduce the total SFR (e.g., Tonnesen \& Bryan 2012).  However,
it has also been also argued (Bekki 2014) that star formation can also
be enhanced by  RP, and that the reduction/enhancement  will depend on
model parameters such as halo mass, peri-centric distance with respect
the the centre of the cluster etc.

In semi-analytic  models (SAMs) of galaxy  formation, the descriptions
of the effect of RP  on the gas component  are very simplistic.  In  
early versions, satellite  galaxies  were assumed  to  lose  their hot  gaseous  haloes
immediately  after falling  into  a bigger  halo  (e.g., Kauffmann  et
al. 1993; Somerville et al. 1999; Cole  et al. 2000; Kang et al. 2005;
Bower et al.  2006; De Lucia \& Blaizot 2007  (hear after DLB07)).  It
was then found  (e.g., Kang \& van  den Bosch 2008; Kimm  et al. 2009)
that  the   instantaneous  stripping  of   the  hot  gas   causes  the
over-prediction of red  satellite galaxies in the  clusters.  In later
models (e.g., Kang  \& van den Bosch 2008; Font  et al. 2008; Weinmann
et al. 2010; Guo et al.  2011, hereafter Guo11)), the stripping of hot
gas is treated in a more continuous way, i.e the mass loss rate of hot
gas halo is assumed  to be the same as that of  dark matter subhalo in
which the satellite resides. Including  only the stripping of hot halo
gas is not physically plausible; the stripping of cold gas in galaxies
is  also needed  to account  for the  observed cold  gas depletion  in
satellites (Fabello et al. 2012; Li  et al. 2012b; Zhang et al. 2013).
Stripping of  cold gas  has been  included in  some models  (Okamoto \&
Nagashima  2003; Lanzoni  et al.  2005), and  has been found  to have 
negligible effect on the colors and SFRs of satellite galaxies.

Because  the  stripping  of  gas  in  the  satellite  depends  on  the
competition between the RP from the hot gas and the gravitation of the
satellite itself,  it is important  to model the local  environment of
satellite  galaxies  accurately.  In  N-body  simulations,  the  local
environment at the  subhalo level will be dependent  on the resolution
of the simulation. In low-resolution simulation, the evolution of subhalo can not  be traced to very low mass due to the number threshold used to identify subhalo, and for given mass a subhalo in low-resolution simulation contains less number of particles, making its identification more difficult.  In the central region of halo the identification of subhalo is more challanging as the background density is high (Onions et al. 2012). We also  note that  the physical  descriptions of  processes such  as gas
cooling,  feedback,  stripping  etc,  are   very  often  tied  to  the
properties of  the subhalo (Springel  et al.  2001; Kang et  al. 2005;
Bower et  al. 2006; Guo  et al.  2011;), and these  prescriptions will
then also be dependent on resolution. This will lead to non-convergent
results between  different resolution simulations, as  found in recent
studies (Fu et al. 2013; Lagos et al. 2014; Guo \& White 2014), and it 
also raises the concern that the high fraction of passive galaxies in 
low-resolution simulation is a consequence of resolution effects.

In this paper, we study the quenched fraction of galaxies in different
environments with  and without the  effects of RP stripping.   We adopt
the  version  of  L-Galaxies  model  described in  Fu  et  al.  (2013,
hereafter Fu13), which is a recent version of the Munich semi-analytic
model that  includes the radial  distribution of molecular  and atomic
gas  in galaxy  disks. This  model  allows us  to model  the cold  gas
stripping as a function of radius in  the galaxy . We improve the Fu13
model by 1)  using a consistent description of  physics for satellites
whose  subhaloes  cannot   be  resolved.   Our  resolution-independent
prescriptions can also be applied to  other SAMs based on merger trees
from N-body simulations. 2) We add a model for the effect of RP stripping
on the  cold gas in  galactic disks.   These improvements allow  us to
model ram pressure stripping of cold  gas at different radii in disks,
and to  study how the environment  can affect the HI,  $\h2$, and star
formation.

This paper  is organized as  follows. In  Section 2, we  first briefly
summarize the L-Galaxies  model and then describe the  changes we make
to Fu13 model. In Section 3, we analyze stellar/gas mass functions and
galaxy  clustering,  and we  compare  our  model results  with  recent
observations of  the properties of  satellite galaxies, such  as their
specific   star  formation   rates and gas   fractions. 
We test the degree to which our new models give convergent
results between two N-body simulations of different resolution, and we
analyze the effect of RP stripping in cluster environments. In Section
4, we summarize our results and discuss possibilities for future work.

\section{The Model}

In this section,  we briefly introduce the N-body  simulations used in
this  work as  well as  the L-Galaxies  models, and  then describe  in
detail  the main  changes  to the  physics made  with  respect to  the
previous models.

\subsection{N-body simulations and L-Galaxies model}

Our work  in this paper  is based  on the Munich  semi-analytic galaxy
formation model, L-Galaxies,  which has been developed  over more than
two  decades (e.g., White \& Frenk 1991; Kauffmann et  al.  1993;  Kauffmann et  al. 1999;
Springel et al.   2001; Croton et al. 2006; De  Lucia \& Blaizot 2007;
Guo  et  al.  2011\&2013;  Fu  et  al.  2010  \&  2013;  Henriques  et
al. 2015).  The L-Galaxies model  has been implemented on  two main N-body
cosmological  simulations: The  Millennium  Simulation (hereafter  MS,
Springel et al.  2005)  and Millennium-II simulation (hereafter MS-II,
Boylan-Kolchin et al. 2009).  The two simulations have the same number
of particles  and cosmology  parameters, but the  MS-II has  1/125 the
volume of MS,  but 125 times higher in mass resolution.   Angulo \& White
(2010) developed a method to  rescale the cosmological parameters from
the WMAP1  to the WMAP7 cosmology.   For MS, the box  size is rescaled
from  $500~\rm{Mpc}~h^{-1}$  to   $521.555~\rm{Mpc}~h^{-1}$,  and  the
particle  mass is  changed from  $8.6\times10^8\rm{M_\odot}~h^{-1}$ to
$1.06\times10^9\rm{M_\odot}~h^{-1}$;  for   MS-II,  the  box   size  is
rescaled          to           $104.311~\rm{Mpc}~h^{-1}$,          and
$8.50\times10^6\rm{M_\odot}~h^{-1}$  for the  particle mass.   In this
paper, we follow  Angulo \& White (2010) and use  the runs appropriate
for    the   WMAP7    cosmology,   with    parameters   as    follows:
$\Omega_\Lambda=0.728,                                    ~\Omega_{\rm
  m}=0.272,~\Omega_{\rm{baryon}}=0.045,~\sigma_8=0.807$ and $h=0.704$.

In  SAMs, galaxies  are assumed  to form  at the  centres of  the dark
matter haloes.  The evolution of  the haloes is followed  using merger
trees  from  the  N-body  cosmological  simulations,  and  the  models
describe the physical  processes relevant to the  baryonic matter, e.g
re-ionization, hot gas cooling and cold gas infall, star formation and
metal production, SN feedback, hot  gas stripping and tidal disruption
in satellites, galaxy mergers, bulge formation, black hole growth, and
AGN feedback.   The detailed descriptions of  these physical processes
can be found in Section 3 of Guo11.

In  the   L-Galaxies  model,   galaxies  are  classified   into  three
types. Type  0 galaxies are  those located at  the centre of  the main
haloes  found  with  a   Friends-of-Friends  (FOF)  algorithm  in  the
simulation outputs. A  Type 0 galaxy is a ``central''  galaxy with its
own hot gaseous halo, and the  hot gas distributes isothermally in the
dark matter  halo. The hot gas  can cool onto the  central galaxy disk
through a ``cooling flow'' or a ``cold flow'', and the cold gas is the
source of star formation.  An instantaneous recycling approximation is
adopted  for mass  return from  evolved stars  and for the injection of
metals into the  ISM; this implies that the massive  stars explode as 
SN  at the time when they form.
\footnote {Yates et  al. 2014 considered
more realistic  model for chemical enrichment  of different elements.}
The SN feedback energy  reheats part of  disk cold  gas into the  hot gaseous
halo of the central galaxy. If the  SN energy is large enough, part of
the hot gaseous  halo can be ejected out of  the dark matter potential
and become  ejected gas. With  the growth  of the
dark matter  halo, the ejected  gas will be reincorporated  back to the  central halo.

Both Type 1 and Type 2  galaxies are regarded as satellite galaxies in
the model.  A Type  1 galaxy is  located at the  center of  a subhalo,
which   is  an   overdensity  within   the  FoF   halo  (Springel   et
al. 2001). The haloes/subhaloes contain at least 20 bound particles 
for both MS and MS-II (Springel et al. 2005, Boylan-Kolchin et al. 2009). 
Boylan-Kolchin et al. (2009) have pointed that above this particle number 
(20) the abundance of subhaloes between two simulations differs only about 
30\%, and with more than 50-100 particles the results agree much better.
Type  1 galaxies have their own hot  gaseous haloes, so  
gas  can cool  onto  these  galaxies.  Cold  gas  reheated  by the  SN
explosions in a Type  1 will be added  to the hot gaseous  halo of its
own subhalo or  the halo of the central Type 0 galaxy,  depending on the distance
between the  Type 1  and the  central object.  A Type  2 galaxy  is an
``orphan  galaxy'', which  no  longer has  an  associated dark  matter
subhalo.  A Type 2 galaxy does not have a hot gaseous halo and thus has no
gas cooling and infall. The supernova reheated cold gas from Type 2 is
added to  its central galaxy,  i.e., the associated  Type 0 or  Type 1
object.

Both  Type  1  and Type  2  galaxies  are  initially  born as  Type  0
objects. They become Type 1 when they fall into a group or cluster and
may  later  become  type  2  after their  dark  matter  subhaloes  are
disrupted by  tidal effects or the subhaloes are not well resolved in 
the low-resolution simulation. Type 1  and 2 galaxies may later  merge into the
central galaxy of their host (sub)halo. In Guo11 \& Fu13, the hot gaseous
halo and ``ejected  reservoir'' of a Type 1 galaxy  can be stripped by
RP when  the force of  RP dominates over  its self-gravity.  A  Type 2
galaxy  can be  disrupted  entirely  by tidal  forces  exerted by  the
central object,  if the  density of  the main  halo through  which the
satellite travels at  peri-centre is larger than  the average baryonic
mass density of the Type 2.

In this paper, we update the semi-analytic models of Fu et al. (2013),
which is  a branch of  the recent  L-Galaxies model.  
Compared with  the previous L-Galaxies  models, Fu13
contains the following two main improvements:

\noindent (i)  Each galaxy  disk is divided  into multiple  rings, and
thus the evolution of the radial distribution of cold gas and stars  can be
traced.

\noindent (ii) A prescription for the conversion of atomic gas into molecular gas is
included,  and the  star  formation  is assumed to be  directly proportional  to  the
local surface density of molecular gas $\Sigma_{\rm SFR}\propto\Sigma_{\h2}$.

These improvements  enable  us to calculate whether the  RP at  a certain radius
in the galaxy  disk is sufficient to remove the gas, and thus to  
trace the depletion of atomic and  molecular gas 
in cluster galaxies.

In this paper, We make two further main changes to the Fu13 model:

\noindent  (1)  we  introduce  an   analytic  method  to  trace  the
mass evolution of unresolved subhaloes once they are not resolved by the simulation. 
This enables us to model the evolution of low-mass satellite galaxies in a resolution-independent way. 

\noindent (2) we include new  prescriptions for ram pressure stripping
of the cold gas;

In the following sections, we describe  our modifications in detail.

\subsection{Tracing galaxies in unresolved subhaloes }

As  discussed  in Section  1,  the  properties of  low-mass  satellite
galaxies predicted by L-galaxies will be resolution dependent, because
the treatment of the physics depends on whether or not the subhalo of
the satellite is  resolved by the simulation (i.e.  whether the galaxy
is Type 1 or Type 2). The detailed issues are the following:
\begin {enumerate}
\item For satellite galaxies,  $M_{\rm vir}$,$V_{\rm vir}$,$R_{\rm
  vir}$ and $V_{\rm max}$ are  fixed at the moment  they first fall 
  into a larger halo, whereas  $M_{\rm  sub}$ is  measured  in  each
  simulation output  until the subhaloes are  disrupted. $M_{\rm sub}$
  for  a Type 2 will be fixed at  the last time when
  it was a Type 1, which  will be strongly dependent on the resolution
  of the simulation. 
\item  The hot gas in a subhalo  is assumed to be
  distributed the  same way  as the  dark matter.   When a  subhalo is
  disrupted and the galaxy becomes a Type 2, it is assumed to lose its
  hot gas and ejected gas  reservoirs immediately.  Thus, gas cooling,
  infall and reincorporation  no longer take place if the  galaxy is a
  Type 2.
\item Only  Type 2  galaxies can  be disrupted by  the tidal  force of
  central galaxies.
\end {enumerate}
These  three assumptions  will  cause  inconsistencies between  models
based on  dark matter simulations with  different resolutions, because
many Type  1 galaxies in a  high resolution simulation will  be Type 2
galaxies in low resolution simulations (e.g. Fig.A1 in Guo11).

In the following section, we incorporate the model of Jiang \& van den
Bosch (2014, hereafter JB14) to trace the evolution of a subhalo after
it is no longer resolved by the  N-body simulation. In this way we can
estimate the key  properties of  unresolved  subhaloes,  such as  $M_{\rm
  sub}$, $V_{max}$,  and treat Type
2s in the same way as Type  1s.  We will then show that this procedure
helps to  alleviate the  resolution-dependent properties  of satellite
galaxies in the models.

\subsubsection{An analytic model for subhalo evolution}\label{chap:esub}

According to JB14, the average mass loss rate of a subhalo depends only
on the  instantaneous mass ratio  of the  subhalo mass $m$  and parent
halo $M$,
\begin{equation}
\label{eq:massratio}
\dot m =  - \varphi \frac{m}{{{\tau _{\rm dyn}}}}{\left( {\frac{m}{M}} \right)^\zeta }
\end{equation}
where $\varphi$ and $\zeta$ are free parameters, $\tau_{\rm dyn}$ is the halo's dynamical time
\begin{equation}
\label{eq:tdy}
\tau_{\rm dyn}(z)=\sqrt{\frac{3\pi}{16G\rho_{\rm crit}(z)}},
\end{equation}
where $\rho_{\rm crit}(z)$ is the critical density at  redshift $z$. 
So we can get the subhalo mass $m(t+\Delta t)$ in a static parent halo at $t+\Delta t$ as:
\begin{equation}
\label{eq:esubmass}
m(t+\Delta t)=m(t)[1+\zeta(\frac{m}{M})^\zeta(\frac{\Delta t}{\tau})]^{-1/\zeta}
\end{equation}
where $\tau=\tau(z)/\varphi$ is the characteristic mass loss time scale at redshift $z$. 
We adjust the parameters $\zeta$ and $\varphi$ so that the distribution of the predicted
distribution of $M_{\rm sub}$ for all  Type 1 galaxies  at $z=0$ matches the
distribution of $M_{\rm sub}$ for Type 1 galaxies measured directly from the
$z=0$  simulation output. The best-fit values we find are $\zeta=0.07$, $\varphi=9.5$. 
Note that our $\varphi$ is much larger than the value of JB14 (their best value is 1.34) and the difference is mainly due to the definition of $\rho$. If we use the same definition as JB14, our best fitted $\varphi$ is about 40\% lower than that of JB14.

JB14  also provides  a formula  to estimate  the $V_{max}$  of subhalo
during the evolution. They find that $V_{max}$ evolves more slowly than the
mass evolution, because $V_{max}$ is mainly  determind by the inner  region of the
subhalo which is not strongly afftected by the tidal force of the host
halo. For simplicity and consistency  with Guo11, we keep the $V_{max}$
for the Type  2 satellite also fixed at the value when it was last a Type 0 object. 
This means that the main change with respect to the Guo11 model occurs after 
the subhaloes have been fully tidally disrupted. So we  only apply the above 
model for  subhalo mass evolution after it is not resolved in the simulation.

\subsubsection{A consistent treatment of the physics of satellite galaxies}

In the  Guo11 and Fu13  models, Type 2s lose  all their hot  gas after
their  subhaloes  are disrupted.  Using  the  methodology outlined  in
Sec.\ref{chap:esub}, we can now estimate the evolution of subhaloes to
arbitrarily low  masses , and thus  Type 2 galaxies will  retain their
own  hot gas  haloes and  lose hot  gas continuously  through stripping
processes in  the same way  as Type 1s. The hot gas is assumed to have 
a isothermal distribution: 
\begin{equation}
\label{eq:rho}
\rho_{hot}(r)=\frac{M_{\rm hot}}{4\pi R_{\rm vir}r^2}
\end{equation}
When  Type 2s fall  within the
virial radius of the central galaxy, we calculate the stripping radius
as $R_{\rm strip}=min(R_{\rm tidal},R_{\rm r.p.})$.  We use Eq. 25 and
26  in Guo  et al.  (2011) to  calculate $R_{\rm  tidal}$ and  $R_{\rm
  r.p.}$:
\begin{equation}
\label{eq:tidal}
R_{\rm tidal}=(\frac{M_{\rm sub}}{M_{\rm sub,infall}})R_{\rm vir,infall}
\end{equation}
where $M_{sub}$ is the subhalo mass given by Eq.1, $M_{\rm sub,infall}$ and $R_{\rm sub,infall}$  are the dark matter mass
and virial radius of the subhalo at the last time when it was Type 0.

\begin{equation}
\label{eq:ramp}
\rho_{\rm sat}(R_{\rm r.p})V^2_{\rm sat}=\rho_{\rm par}(R)V^2_{\rm orbit}
\end{equation}
where $\rho_{\rm sat}(R_{\rm r.p})$  is the hot  gas density  of the satellite at
$R_{\rm r.p}$; $V_{\rm sat}$ is the virial velocity of the subhalo;
$\rho_{\rm par}(R)$  is the  hot  gas  density of  the  main  halo at  the
distance $R$; and $V_{\rm orbit}$ is the orbital velocity of the satellite
(we simply use the virial velocity of the main halo).

All the hot gas beyond $R_{\rm strip}$ is removed and added to the hot
gas component of  the parent central galaxy.  Then we  set the hot gas
radius to be  $r_{\rm hot}=R_{\rm strip}$. The hot gas  in Type 2s will
cool and fall onto the galaxy disk with an exponential surface density
distribution
\begin{equation}
\label{eq:gasprofile}
\Sigma_{\rm gas}(r)=\Sigma^{(0)}_{\rm gas}\exp (-r/r_{\rm infall})
\end{equation}
where the  infall  scale  length
$r_{\rm infall}=(\lambda/\sqrt{2})r_{\rm vir}$.  We keep the original spin
parameter $\lambda$ and $r_{\rm vir}$  when the galaxy was last a Type 0  unchanged.

In the models of  Guo11 and Fu13, SN feedback reheats  the cold gas in
the disk,  and if there is  remaining energy, hot gas  will be ejected
out of  the halo.  In the Guo11 code, the supernova  reheating efficiency
$\epsilon_{\rm disk}$ is written as
\begin{equation}
\label{eq:guo19}
\epsilon_{\rm disk}=\epsilon \times [0.5+(\frac{V_{\rm max}}{V_{\rm reheat}})^{-\beta_1}]
\end{equation}
The  supernova ejection  efficiency is written as
\begin{equation}
\label{eq:guo21}
\epsilon_{\rm halo}=\eta \times [0.5+(\frac{V_{\rm max}}{V_{\rm eject}})^{-\beta_2}],
\end{equation}
The efficiency  of the  SN feedback  in Type  2s is
assumed to scale with the maximum circular velocity $V_{\rm max}$ of 
its central galaxy and reheated gas from Type 2s is added to the hot gas component
of the central galaxy. We note that in Guo11, Type 2s have 
no hot gas component.  
Here, we assume SN  in Type 2s will reheat cold  gas to its own
hot  gas component and allow for an ejected gas reservoir 
in Type 2s in the same way as for Type 0s and 1s. Due to the fact that hot gas 
is stripped from satellites, only a fraction $R_{\rm hot}/R_{\rm vir}$ of 
reheated gas remains in the subhalo and the rest is returned to the main halo. 
We replace $V_{\rm max}$ of the Type 2s' central galaxy with 
the $V_{\rm max}$ for the Type 2 galaxy is taken as the value of $V_{\rm max}$ 
when it was last a Type 0.  Note that in our new model, 
the SN heating efficiency in a Type 2 is determined by its own $V_{\rm max}$, 
which has a lower value than the $V_{\rm max}$ of the central galaxy that 
is used to scale the SN reheating efficiency in a satellite galaxy in the 
Guo11 and Fu13 models. This choice is also motived by some recent work 
(e.g., Lagos et al. 2013; Kang 2014) which shows that SN feedback is more 
likely determined by the local galaxy potential. Usually the $V_{\rm max}$ of satellite 
is lower than $V_{\rm max}$ of the central, so the SN heating efficiency is 
higher in satellite galaxies in our model and we will later show in Section 3.2 that 
it leads to a slightly better agreement with the observed galaxy two-point correlation 
function on small scales in low stellar mass bins.

The hot gas  in the subhaloes of  Type 2s will cool and  fall onto the
gas disks of Type  2s later on.  With the growth  of the dark matter halo,
the ejected  gas in the subhalo  will be  reincorporated into the  hot gas
again.  The reincorporation efficiency is
\begin{equation}
\label{eq:guo23}
\dot{M}_{\rm ejec}=-\gamma(\frac{V_{\rm vir}}{220})(\frac{M_{\rm ejec}}{t_{\rm dyn,h}}),
\end{equation}
where $\gamma$ is free  parameter, $t_{\rm dyn,h}=R_{\rm vir}/V_{\rm vir}$ is the halo
dynamical  time. 

Finally, we  note that in  the Guo11 model,  only Type 2  galaxies are
disrupted  by tidal  forces (see  Sec. 3.6.2  in Guo11).   In our  new
model, we follow the evolution  of subhaloes analytically, so we treat
the tidal disruption for  Type 1s and 2s in the  same way. A satellite
galaxy (Type 1 or Type 2) will  be disrupted by the tidal force of the
main  halo, when  1)  it has  more baryonic  matter  than dark  matter
($M_{\rm bar}>M_{\rm sub}$);  2) its average baryonic  mass density is
lower than main halo density at the peri-centre of its orbit.

\begin{figure*}
\centering
\includegraphics[scale=1.1]{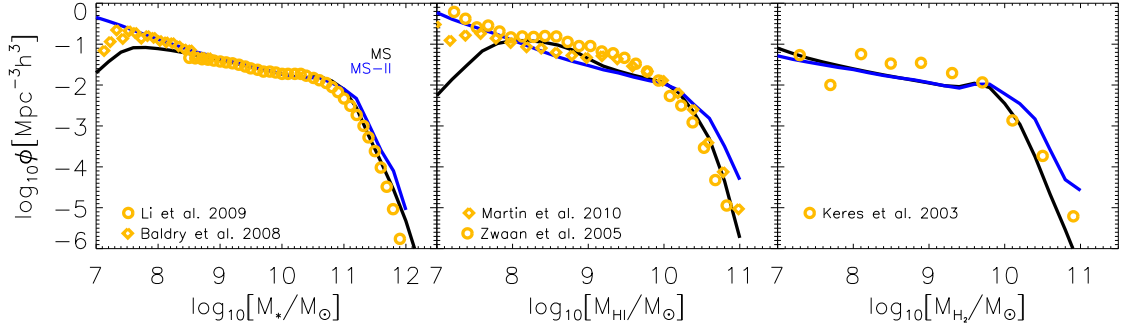}
\caption{The stellar,  HI and $\h2$ mass functions at  $z=0$ from
the new model compared with  observations. The solid curves show results for  
the  new   model  without the process  of ram  pressure 
stripping  of cold  gas. Blue and black curves show results for generated from the 
MS   and  MS-II simulations, respectively.
For the observations,  stellar mass functions are from
Li \& White  (2009) and Baldry et al. (2008); the  HI mass functions are
from Zwaan  et al. (2005) and  Martin et al. (2010)a the;  $\h2$ mass
function is from Keres et al. (2003).}
\label{fig:mfv1t}
\end{figure*}

\subsection{The physical prescriptions for ram pressure stripping of cold gas}

In  the L-Galaxies model,  ram pressure  strips  
only  the hot  gas component in the  satellite galaxies, while the cold  gas component in
the ISM is  not affected.  
Following the prescriptions of Gunn  \& Gott  (1972),  we consider  a satellite  galaxy
moving through  its host halo.  The RP  force can be  written as,
\begin{equation}
\label{eq:Prp}
P_{\rm r.p}(R)=\rho_{\rm ICM}(R)v^2
\end{equation}
where  $v$ is  the  orbital  velocity of  the  satellite,  which we take to be the 
virial velocity  of its parent halo. $\rho_{\rm ICM}$
is the  density of  the hot  gas of  the parent  halo as Eq.(\ref{eq:rho}). 
When the ram  pressure exceeds  the interstellar pressure  $P_{\rm ISM}$, cold
gas will  be stripped. We adopt the Eq.(\ref{eq:Pism})  given by
Tecce et al.(2010):
\begin{equation}
\label{eq:Pism}
P_{\rm ISM}=2\pi G\Sigma_{\rm disc}(r)\Sigma_{\rm gas}(r)
\end{equation}
where $r$ is the radius to the centre of satellite. $\Sigma_{\rm disc}$ is
the surface density of the galactic  disc, which equals to the sum of the 
cold gas and stellar surface densities:
\begin{equation}
\label{eq:sigmas}
\Sigma_{\rm disc}=\Sigma_{*}(r)+\Sigma_{\rm gas}(r)
\end{equation}

We calculate $P_{\rm ISM}$ in each radial concentric ring in satellite
galaxies  based  on the  division  of  the  disk into  multiple  rings
introduced  in the  Fu13  model.  In  our  model, the  cold  gas in  a
satellite galaxy can  be stripped by RP only when  the satellite falls
within  the virial  radius $R_{\rm  vir}$  of the  central galaxy  and
$P_{\rm  r.p}(R)\geq P_{\rm  ISM}(r)$. The  stripped cold  gas of  the
satellite  is  added   to  the  hot  gas  component   of  its  central
galaxy.  According  to  Eq.  (\ref{eq:Prp})  \&  (\ref{eq:Pism}),  the
criterion for RP stripping for a  satellite galaxy at the distance $R$
to the centre of its parent halo is written as:
\begin{equation}
\label{eq:Ri}
2\pi G\left[ {{\Sigma _*}(r) + {\Sigma _{{\rm{gas}}}}(r)} \right]{\Sigma _{{\rm{gas}}}}\left( r \right) \le {\rho _{{\rm{ICM}}}}\left( R \right)v^2
\end{equation}
From the  radial distribution  of cold  gas $\Sigma_{\rm  gas}(r)$ and
stellar surface density $\Sigma_{*}(r)$, we evaluate the stripping   radius $r$ in
Eq. (\ref{eq:Ri}) and assume that cold gas exterior to this radius $r$ will be stripped .

The above description  is very simplistic. It is not  clear if all the
cold gas will be stripped immediately when the RP force is larger than
the  gravity  of  the  satellite  itself.  To  take  account  of  this
uncertainty,  we define  a  stripping fraction  $f_{\rm  rps}$ as  the
fraction of the cold  gas stripped by ram pressure in  the region of a
satellite galaxy  where $P_{\rm  r.p}(R)\geq P_{\rm ISM}(r)$.   In the
simplest case, $f_{\rm rps}$ is $100\%$,  which means all the cold gas
in  the  region where  $P_{\rm  r.p}(R)\geq  P_{\rm ISM}(r)$  will  be
stripped by RP. This simple assumption  causes a sudden cut off of the
cold gas  radial profile, i.e  the cold  gas radial profile  becomes a
step function at  the radius $r$. To ensure  a continuously decreasing
cold gas radial  profile, we modify the stripping law  as follows : if
$P_{\rm r.p}\geq P_{\rm  ISM} $, the stripping ratio  $f_{\rm rps}$ is
related to the difference between $P_{\rm r.p}$ and $P_{\rm ISM}$ as:
\begin{equation}
\label{eq:caseB}
f_{\rm rps}=\begin{cases}
0,& P_{\rm r.p}<P_{\rm ISM}\\
\frac{P_{\rm r.p}-P_{\rm ISM}}{P_{\rm r.p}},& P_{\rm r.p}\geq P_{\rm ISM}
\end{cases}
\end{equation}
The guarantees that only part of the cold gas $M_{\rm stripped}=f_{\rm rps}M_{\rm cold
  gas}(r)$  in the  region where  $P_{\rm r.p}(R)\geq  P_{\rm ISM}(r)$
will be stripped.

\begin{figure*}
\centering
\includegraphics[scale=0.9]{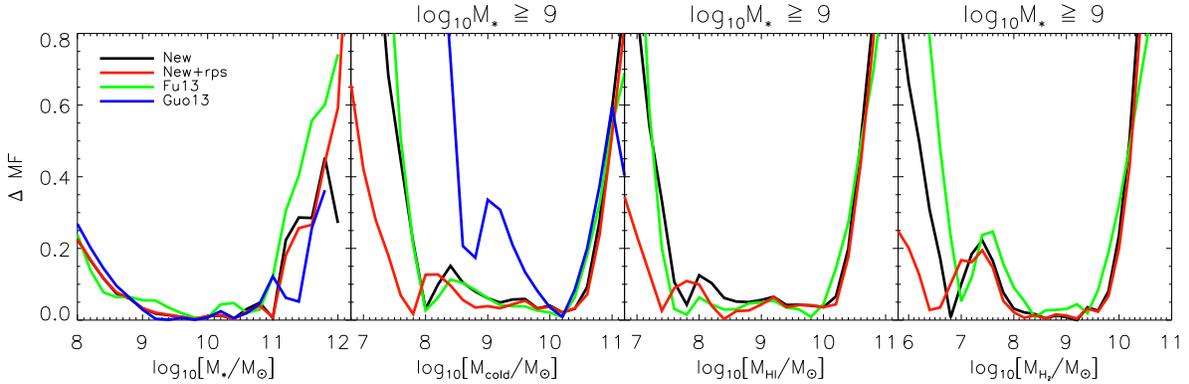}
\caption{Convergence check: the difference  of the predicted stellar, cold
gas, HI  and $\h2$ mass  functions between  MS and MS-II.  
The black/red lines are for our new model without/with ram pressure 
stripping of cold gas. The green/blue lines are for the Fu13 and Guo13 models.}
\label{fig:mfserr}
\end{figure*}

\section{Performance of the new model}

In  this section,  we compare  our new  model with the Fu13 model and
explore if we obtain more convergent and/or improved results for the following
galaxy properties: the stellar/gas mass function and the two-point correlation function.
These two quantities are often used as basic tests of the overall model. 
We then  study the fraction of quenched galaxies as a function of halo mass
and compare with observational data. In the following subsections, we abbreviate our 
model as ``new'' if the option for cold gas stripping is switched off, and it is 
labeled as ``new +rps'' if the cold gas stripping is turned on.

\subsection{The stellar and cold gas mass functions}

In Fig. \ref{fig:mfv1t}, we plot the model mass functions of stars, HI
and $\h2$  at $z=0$  compared with the  observations. The  ``new'' model
results (with  RP stripping of  cold gas off)  are shown as  black and
blue lines  for the MS  and MS-II simulations, respectively.   All the
free parameters  of the  models are  fixed as in  Fu13 (see  the Table
1.  in that  paper),  with  the exception  of  the  hot gas  accretion
efficiency  onto  black  holes  $\kappa_{\rm AGN}$ which is changed from  $1.5  \times
10^{-5}\rm{M_\odot} \rm yr^{-1}$ to  $1.5 \times 10^{-6}\rm{M_\odot} \rm
yr^{-1}$. Tuning  this parameter is  necessary in order to  better fit
the stellar mass function  at the high mass end. As  can be seen, with
this minor change, our new models  are in very good agreement with the
observational data at $z=0$ for both simulations. 

Fig. \ref{fig:mfv1t} also shows that there is difference in the predicted mass 
functions at the low-mass end between the two simulations.  
In Fig.  \ref{fig:mfserr}, we check the divergence between two simulations 
in detail, where the difference defined as
$\Delta  \rm MF=|\rm  \log_{10}MF_{MS}-\log_{10}MF_{MS-II}|$. 
The left panel shows that for low-mass galaxies ($\rm log_{10} M_∗ < 8.5$) 
there is obvious difference in the mass function, and all the models 
have similar predictions (also see Fig.7 in Guo11). 
Such a difference is expected and it is basically due to low-mass central 
galaxies not being resolved in the low-resolution simulation.
In fact the simulation resolution can have more complicated effects on 
the galaxy properties. For example, if the merger trees of some massive 
halos in the MS simulation are not well-resolved at higher redshifts, 
this will lead to divergence in the properties of both the central and 
the satellite galaxies in those halos. Overall we find that the stellar 
mass function is more convergent for galaxies with $\rm log_{10}M_{*} > 8.5$ 
and in our later analysis we only focus on these galaxies.

\begin{figure*}
\centering
\includegraphics[scale=1.0]{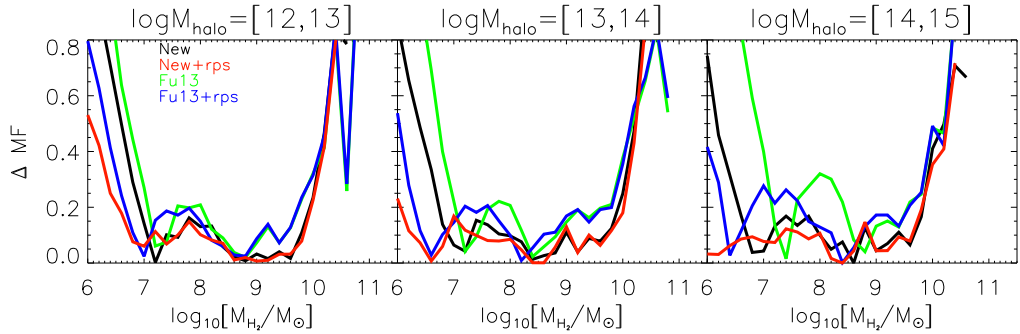}
\caption{Convergence check: the difference of the predicted $\h2$ mass
  functions of  the satellites ($\rm  log M_*>9.5$) in  different halo
  mass bins between MS and MS-II.  The black/red lines are for our new
  model  without/with   ram  pressure  stripping  of   cold  gas.  The
  green/blue lines are for the  Fu13 and Fu13+rps models. More obvious
  improvement in convergence is seen in massive haloes.}
\label{fig:h2err}
\end{figure*}

The right panels of Fig. \ref{fig:mfserr} show the convergence test on
the  cold  gas  mass  functions,   also  divided  into  HI  and  $\h2$
components. The sample is selected with $\rm log_{10}M_*>9$.
It  is found that, compared  to the Guo11  model, our new
model produces more  convergent results for the cold  gas mass. In the
Guo11  model,  there  is  a  threshold  in  cold  gas  mass  for  star
formation which implies that all satellite galaxies contain cold gas 
even when they stop forming stars.  In a  low-resolution simulation,  the satellite  will soon
become  type 2  and  gas cooling  will stop  as  the hot  halo gas  is
immedaitely  stripped.   So  there  will   be  more  cold  gas   in  the
low-resolution run for the Guo11  model. Compared to the Fu13 model on
which  our model is  based, it  is seen  that our  new model  does not
produce a more convergent results on the cold gas and HI gas mass, but
we obtain slightly  better convergence on the $\h2$ mass (right
panel), and the  convergence is more obvious in  our new+rps model. 

  In Fig.   \ref{fig:h2err} we further check  the convergence in
  different halo mass bins. It is clearly seen that the convergence in
  the   new  model   is  better   at  $10^{7}M_{\odot}   <  M_{H2}   <
  10^{10}M_{\odot}$  than  the Fu13  model.  The  new+rps will  mostly
  affects the low-mass end ($<10^{7}M_{\odot}$) as rps is efficient to
  strip the cold gas if there is  little of it. In the Fu13+rps model,
  the H2 mass function is converged  only at the low-mass end, but not
  at  the  high-mass end.  This  plot  clearly  shows that  without  a
  sub-resolution treatment of the star formation physics (such as Fu13
  model),  it is  difficult  to achieve  convergence  in high-mass  H2
  galaxies  and  rps  is  effective in  convergence  in  low  gas-mass
  galaxies.   As in  the Fu13  and our  model, the  star formation  is
  determined by local $\h2$ gas density, it is expected that our model
  will produce more convergent results on the fraction of passive/star
  forming    galaxies,   and    this   is    shown   in    detail   in
  Sec.\ref{chap:quech}.

\subsection{The projected two-point correlation functions (2PCF)}

\begin{figure*}
\centering
\includegraphics[scale=1.0]{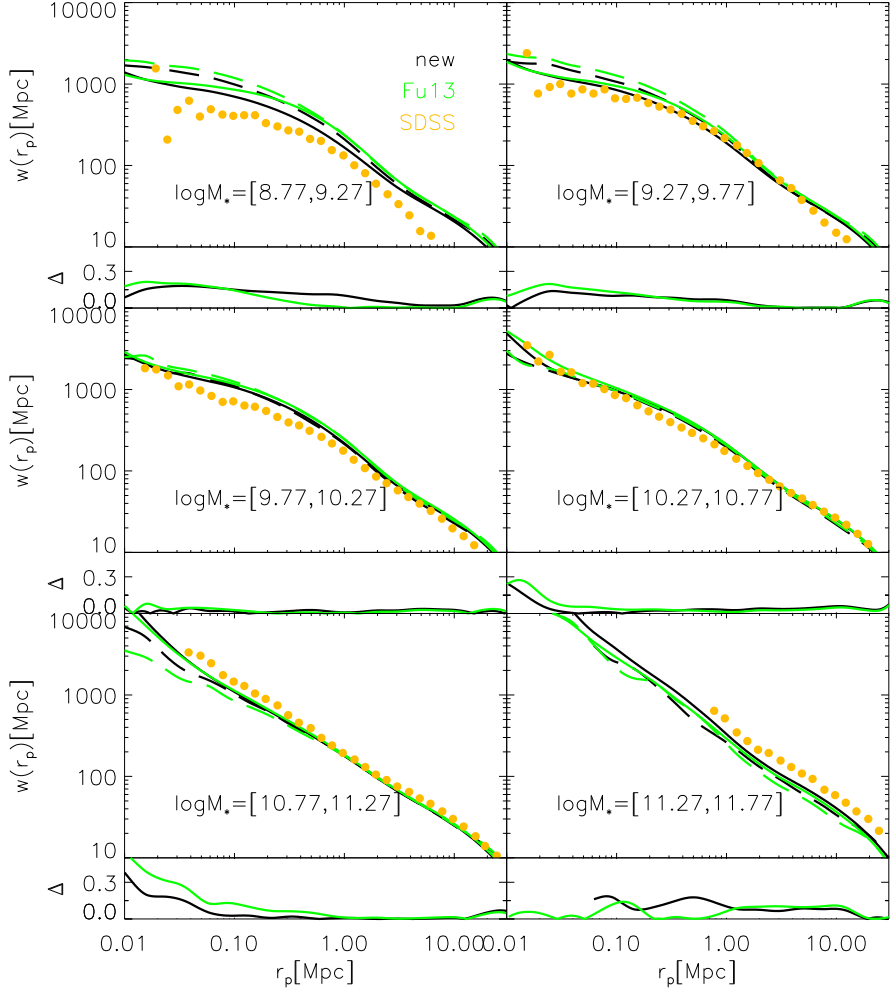}
\caption{The projected 2PCFs for  galaxies in different  stellar mass
bins (top graph in each panel)  and absolute values of difference of the amplitude of the 2pcfs between MS and MS-II, 
$\Delta=|\log_{10}w_{\rm MS}-\log_{10}w_{\rm MS-II}|$ (lower graph in each panel). 
The results from MS are shown with solid lines, and those from 
MS-II are shown using dashed lines. Black curves show results for our new models
and green curves show results for the Fu13 models. The orange dots are data from SDSS.
 Here we only show the results from our new model without RP stripping of cold gas.}
\label{fig:2pcf}
\end{figure*}

The two-point correlation function  (2PCF) is another important galaxy
statistic, because  it describes how  galaxies are distributed  in and
between dark  matter haloes.  On small  scales, the  clustering depends
strongly on how satellites are distributed in massive haloes, so it is
interesting to  ask if the  model results are dependent  on simulation
resolution.  Another  reason to check  the 2PCF  is that, as  shown by
Kang (2014), strong SN feedback in satellites can decrease the 2PCF on
small scales,  and this results  in better agreement with  the observations.   Our new
model  also  includes  a  feedback  model similar  to  Kang  (2014)  for
satellite galaxies,  in which the  feedback efficiency depends  on the
local gravitational potential of the satellite.  It is thus worthwhile
to check if our new model  can give better agreement with the observed
2PCFs.

In Fig.  \ref{fig:2pcf}, we plot the projected galaxy 2PCFs for the Fu13 model in green and
for our new model in black. Results are plotted for both the MS and MS-II simulations
and the difference between the two is shown as in inset at the bottom of each panel.
The different panels show results for  galaxies in different bins of stellar mass.  
The data points are from Li et al. (2006) and are computed using the large-scale
structure sample from the  SDSS DR7. Here we do not show the results from the ``new+rps'' --
the results are very similar to the model without cold gas stripping.

\begin{figure*}
\centering
\includegraphics[scale=1.0]{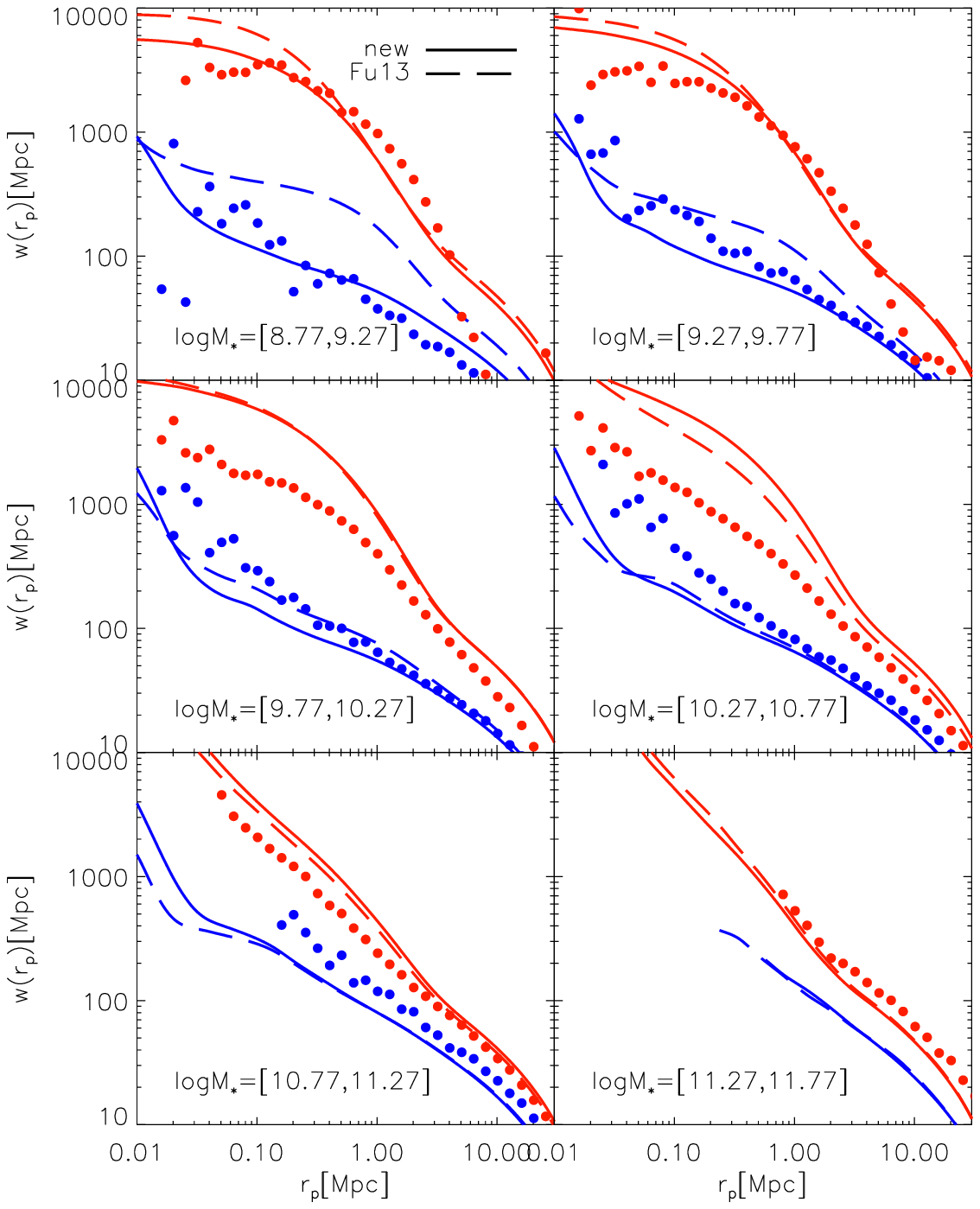}
\caption{As in Fig. \ref{fig:2pcf}, but separately for  red and blue galaxies.}
\label{fig:2pcfr}
\end{figure*}

Fig.~\ref{fig:2pcf} shows that our new model produces better agreement
with the  observed 2PCFs  compared to the  Fu13 model,  especially for
low-mass galaxies. As discussed in the 
previous  section, our new model employees a
feedback prescription  where the SN ejection  efficiency is dependent
on $V_{max}$ of the subhalo, and  not the central halo, as in previous
models. So the  feedback efficiency in satellites in our  new model is
 larger, and this decreases the mass growth of satellites
after infall  with respect to  the results of  Fu13. As shown  by Kang
(2014), strong  feedback in  satellites flattens the  satellite galaxy
mass  function  in  massive   haloes,  thus  reducing  the  clustering
amplitude on small scales.

The  panel inserts  show that  the disparity  in clustering  amplitude
between the  MS and  MS-II simulations  is slightly  lower in  the new
model compared to Fu13 model. The new model shows a little ``better''
convergence at small scales, but in some mass bins new model get ``worse''
at $\rm \sim Mpc$ scales. However, these difference on 2PCFs between
the  two simulations  is  always  small (often  lower  than 0.2  dex),
indicating  that  resolution  is  not of  primary  importance  in  the
prediction of galaxy clustering.

In Fig. \ref{fig:2pcfr} we show 2PCFs for galaxies classified into red
and blue  according to their $g-r$  colors as described in  Guo11. The
red/blue lines  are model predictions  for red/blue galaxies,  and the
red/blue points  are data points from  Li et al. (2006).   As shown in
the  previous  figure,  the  difference   between  the  MS  and  MS-II
simulations are small for 2PCFs, so  we only show results from the MS.
The new model  fits better the color-dependent  clustering in low-mass
binsi, particularly for blue galaxies in the low mass bins.   In  
intermediate  mass   bins  ($\log  M_*=[9.77,10.77]$)  the
clustering  of red  galaxies  is  too strong,  and  too  low for  blue
galaxies for both models. This is because both the Fu13
and our models over-predict the fraction of red satellite galaxies and
under-predict   the   fraction   of   red   centrals,   as   seen   in
Fig.\ref{fig:ssfr1}. Both models also fit  well at the highest stellar
masses.

\subsection{Satellite quenching and cold gas depletion}

In this subsection, we will investigate the effects of RP stripping of
cold  gas  in galaxies.   We  will  study how  ram-pressure  stripping
changes  the quenched  fraction of  galaxies as  a function  of galaxy
mass,  halo mass  and cluster-centric  radius, and  we will  also show
comparisons with recent observations.  As in the previous subsections,
we will also address the issue  of convergence by showing results from
both the MS and the MS-II simulations.

\subsubsection{Where is ram-pressure stripping most effective?} \label{chap:ram}

To understand which galaxies have been affected most by ram-pressure stripping, 
we define the cumulative stripped cold gas fraction  as
\begin{equation}
\label{eq:fsp}
f_{\rm sp}=\frac{M_{\rm asp}}{M_{\rm asp}+M_{\rm cold gas}}
\end{equation}
where  $M_{\rm asp}$  is the  cumulative mass  of stripped  cold gas
throughout the formation history of  a galaxy ( evaluated by summing up the stripped
cold gas mass in the main progenitor), and  $M_{\rm cold gas}$ is  its current
cold  gas  mass.  We focus  on
satellite galaxies in  rich groups and  clusters,  and  select those with
$M_*>10^{9}\rm{M_\odot}$ in haloes with mass  $M_{\rm halo}>10^{13}\rm{M_\odot}$.

\begin{figure}
\centering
\includegraphics[scale=0.8]{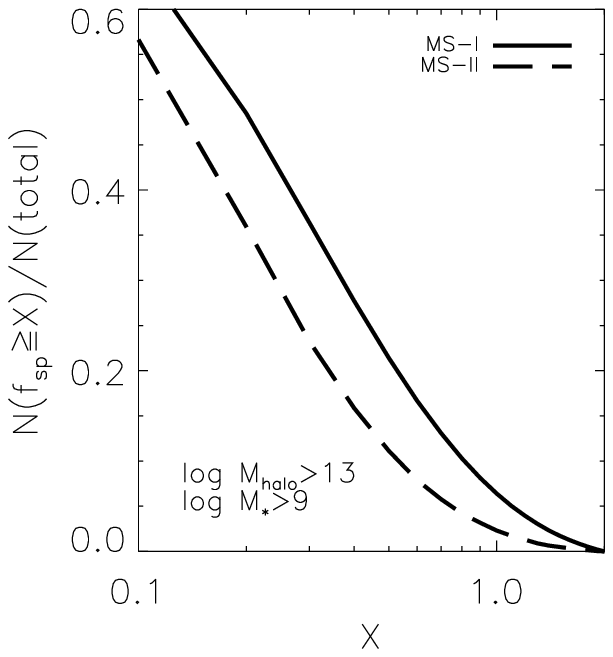}
\caption{The fraction of galaxies  with stripping fraction $f_{\rm sp}\geq
  x$ as a function of $x$. Solid  and dashed lines are based on MS and
  MS-II haloes respectively. Here galaxies are selected with stellar
  mass        $M_*>10^{9}\rm{M_\odot}$        in       haloes        with
  $M_{\rm halo}>10^{13}\rm{M_\odot}$.}
\label{fig:sphist}
\end{figure}

\begin{figure}
\centering
\includegraphics[scale=0.8]{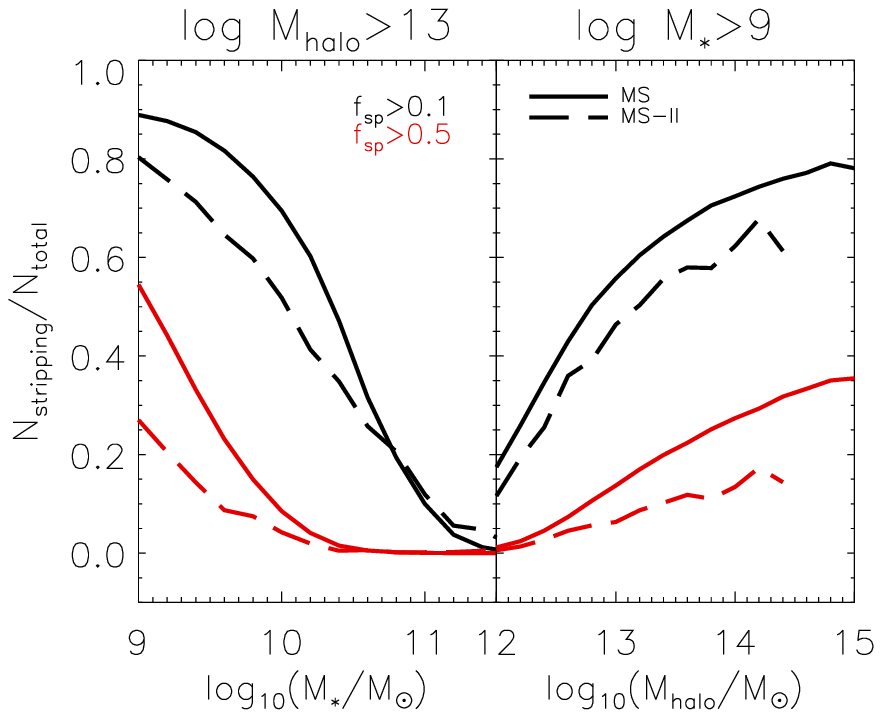}
\caption{The  number fraction  of  galaxies  with $f_{\rm sp}>0.1$  (black
  curves) and $f_{\rm sp}>0.5$ (red curves)  as a function of stellar mass
  and halo  mass. Solid  and dashed  lines are based  on MS  and MS-II
  haloes respectively.}
\label{fig:sp}
\end{figure}

Fig.\ref{fig:sphist} shows the fraction  of galaxies with $f_{\rm sp}$
larger than a  certain value $x$ ($0.1<x<1.0$). Results  are shown for
both MS and MS-II haloes and are seen to agree to within $\sim 10 \%$.
We find that about 50\% galaxies in massive haloes have $f_{\rm sp}\geq
0.1$.  We then define a galaxy with $f_{\rm sp}\geq 0.1$ as having had
significant  cold gas  stripping, and  we plot  the fraction  of these
galaxies  $N_{\rm stripping}/N_{\rm  total}$ as  functions of  stellar
mass and halo  mass in Fig.\ref{fig:sp}. As can be  seen, the fraction
of  galaxies with  significant stripping  increases steeply  with halo
mass, but decreases with stellar mass.

If we increase the significant stripping threshold from 0.1 to 0.5, we
find  that  the stripped  fraction  in  haloes of  $10^{13}  M_{\odot}$
decreases from 0.6 to 0.2.  Br{\"u}ggen  \& De Lucia (2008) found that
about   one  quarter   of  galaxies   in  massive   clusters  ($M_{\rm
  halo}>10^{14}\rm{M_\odot}$)  are subjected  to  strong ram  pressure
that causes loss of all gas and more than 64\% of galaxies that reside
in a  cluster today have  lost substantial  gas. Our new  model agrees
well  with  these  results.   We  also  conclude  that  most  low-mass
satellite galaxies in massive  clusters will lose significant fraction
of  their cold  gas by  RP  stripping, and  some will  lose all  their
interstellar cold gas.

\subsubsection{ Effect of ram-pressure stripping on the quenched fraction of satellite galaxies}\label{chap:quech}

\begin{figure*}
\center
\includegraphics[scale=0.8]{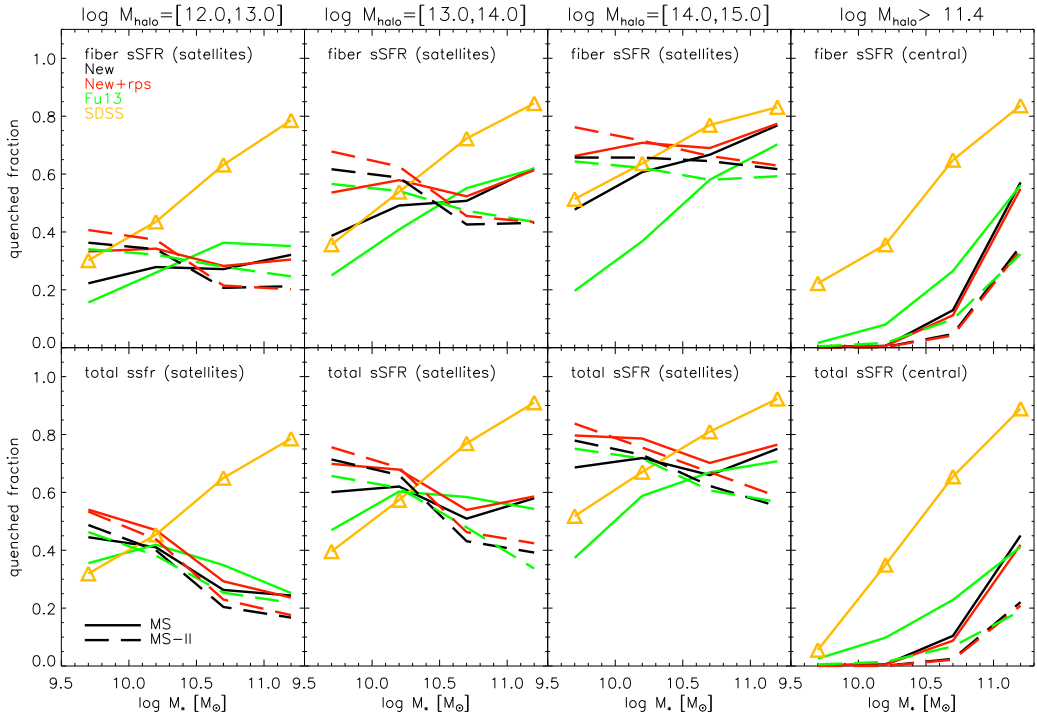}
\caption{The quenched  fraction of galaxies  as a function  of stellar
  mass: the left three columns are for satellite galaxies in different
  halo mass bins, and the right column is for central galaxies in all
  haloes  with  masses greater  than  $10^{11.4}  M_{\odot}$.  The  top
  panels show results  when quenched galaxies are  defined using sSFRs
  measured within  the SDSS fiber,  and the lower panels  show results
  when corrected total star formation  rates are used instead. In each
  panel, the  triangles are the SDSS  data and the colored  lines are
  different  model  predictions  (solid   for  MS,  dashed  for  MS-II
  results).}
\label{fig:ssfr1}
\end{figure*}

The fraction of  quenched galaxies is known to  depend on environment.
It has also been shown (e.g., Weinmann et al. 2006) that early versions
of SAMs (e.g., Croton et al.  2006) over-predicted the fraction of red
satellites in  all environments.  It was suggested  that instantaneous
stripping  of  the  hot  halo  gas of  satellites  was  to  blame.  By
introducing  a non-instantaneous  stripping model,  later SAMS  (e.g.,
Kang \&  van den Bosch  2008; Font et al.  2008) were able  to produce
more blue galaxies, but the  agreement was still not satisfactory.  In
the Guo11 and Fu13 models, the stripping of hot halo gas is modeled as
a gradual process.   Recently Henriques et al. (2015)  showed that the
latest version  of L-Galaxies  was able to  reproduce the  overall red
galaxy fraction,  but they did not  analyze central/satellite galaxies
separately, nor  did they  examine the dependence  of red  fraction on
halo mass and radial distance from the centres of groups and clusters.
 
Wetzel et al. (2012) re-analyzed  the fraction of quenched galaxies in
groups and clusters  using SDSS DR7 data. They  classified galaxies as
quenched  based on  their specific  star formation  rates rather  than
their  colors. This  permits a  more direct  comparison with  models,
because the galaxy color is  a complicated function of star formation
history, metallicity and  dust extinction.  The analysis  of Wetzel et
al. (2012) made use of global star formation rates with fibre aperture
correction as given  in Salim et al. (2007). These  corrections may be
prone to  systematic effects, because there  can be a large  spread in
broad-band colors at a given specific star formation rate and because
the corrections  are calibrated using average  relations between these
two quantities.   The corrections also  account for two thirds  of the
total star formation rate on average.  In contrast, the star formation
rate measured inside  the fiber aperture is derived  directly from the
dust-corrected H$\alpha$  luminosity and  should be more  reliable. In
the following comparison, we present the results on quenched fractions
using both total and fiber  aperture specific star formation rates. We
also  examine  the  radial  dependence of  the  quenched  fraction  in
different halo mass bins.

Following the  same procedure  described in Wetzel  et al.  (2012), we
extract   galaxies  from   the   MPA-JHU  SDSS   DR7  catalogue   with
$M_{*}>10^{9.5}\rm{M_\odot}$          at         $z<0.04$          and
$M_{*}>10^{9.5}\rm{M_\odot}$ at $z=0.04\sim0.06$  that are included in
the group catalogue of Yang et  al.(2007).  We use the stellar masses,
the fibre sSFRs  (specific star formation rates) and  total sSFRs from
the MPA-JHU  SDSS DR7  database.  Following Wetzel  et al.,  we define
galaxies  with   $sSFR<10^{-11}yr^{-1}$  as  quenched   galaxies,  and
calculate  the quenched  fraction of  satellite galaxies  in different
bins of halo mass, stellar  mass and cluster-centric radius, scaled to
the virial  radius of  the halo.  We  find that  $f^{sat}_Q$ increases
with the stellar mass of the satellite, host halo mass and distance to
the center  of the halo  and our  results are largely  consistent with
those presented  in Wetzel et  al.(2012) (see Fig.  \ref{fig:ssfr1} to
\ref{fig:dssfr1}).

In the models, we select galaxies with $M_{*}>10^{9.5}\rm{M_\odot}$ at
$z=0$  from  the MS  and  MS-II  simulations.   We  use the  FOF  halo
properties  given  by  the  N-body   simulation  to  define  the  halo
mass.  However, in  order to  account  for projection  effects in  the
simulation,  we  convert  3D  distances  to  2D  distances  by  simply
projecting  the galaxies  onto the  X-Y plane  in the  simulations and
selecting galaxies within $\Delta V_z\leq 500km/s$ to the halo central
as group  members.  This is  a reasonably close representation  of the
Yang  et al.  (2007) procedure  to  select galaxy  groups in  redshift
space.

In Fig. \ref{fig:ssfr1} and Fig. \ref{fig:ssfr2}, we show the quenched
fraction of satellite galaxies as function  as their stellar mass in a
set of different halo  mass bins, and as a function  of host halo mass
in a set  of different stellar mass bins. In  Fig. \ref{fig:ssfr1}, we
also show the quenched fraction of  central galaxies in all haloes with
$\log M_{\rm halo}>11.4$  in the right  panel. In both  figures, results
from our new model without  ram-pressure stripping are shown in black,
from our new  model with cold gas  stripping in red, and  from Fu13 in
green. The solid  lines are results from the MS,  and the dashed lines
are  for the  MS-II simulation.  The  yellow triangles  show the  SDSS
results.  In each  figure, we plot the results for  fiber sSFRs in the
upper panels, and  for total sSFRs in the lower  panels. At our median
redshift, the 3 arcsec aperture of the fiber corresponds to a physical
aperture size of $1.5 kpc/h$.

One important conclusion from these two figures is that our new models
(w/o RP  stripping) produce much  more convergent results  between the
two  simulations  than  the  Fu13  model,  particularly  for  low-mass
satellite galaxies.  For example,  in the upper  middle two  panels of
Fig. \ref{fig:ssfr1}  the predicted  quenched fractions at  low stellar
masses differ by factors of 2-3 between  the MS and MS-II for the Fu13
models, but improved to within a factor of 1.5 in our new models. In 
the $\rm log M_{halo}=[14,15]$ panel, as we discussed in Sec. 3.1, the 
convergence is more obviously in the new model. 
The same is seen in
the leftmost panel of Fig. \ref{fig:ssfr2} and for the highest halo 
mass bins in Fig. \ref{fig:dssfr1}. 

We also conclude that the predictions  using the fiber and  total sSFR are
qualitatively very similar. There are significant discrepancies between all the models and
the observations which are much larger than any of the systematics in the way we 
choose to define the boundary between quenched and actively star-forming galaxies.  
The clear {\em qualitative}  discrepancies are the following:
\begin {enumerate}
\item The rightmost panel of Fig. \ref{fig:ssfr1} shows that the quenched fraction 
of central galaxies is lower than the data, across all stellar masses. 
This under-prediction of red centrals agrees with 
previous results based on galaxy color (e.g., Weinmann et al. 2006; Kang et al. 2006). 

\item Comparing Fig. \ref{fig:ssfr1} and \ref{fig:ssfr2}, we see that in the data, the quenched fraction depends on both
stellar mass and on halo mass. At fixed halo mass, there is still a strong dependence of
quenched fraction on the stellar mass of the galaxy. At fixed stellar mass, there is a 
significantly weaker dependence of quenched fraction on halo mass. This indicates that
halo mass has {\em secondary} influence on the star formation histories of galaxies
compared to stellar mass.

\begin{figure*}
\begin{center}
\includegraphics[scale=0.8]{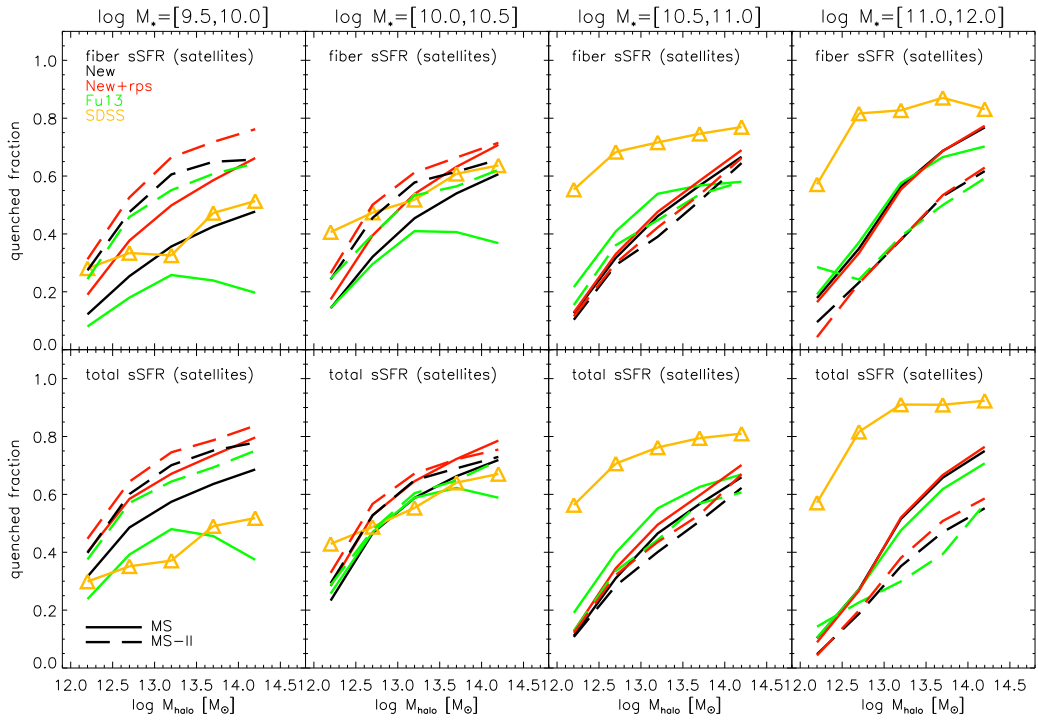}
\end{center}
\caption{As in Fig. \ref{fig:ssfr1}, but now quenched fractions are plotted as a function of
halo mass  in different stellar bins.}
\label{fig:ssfr2}
\end{figure*}

In the models, the {\em opposite} is true. At fixed halo mass, there is generally only a weak
dependence of the quenched fraction on stellar mass. At fixed stellar mass, the dependence of
the quenched fraction on halo mass is very strong. This indicates that in the models,
the influence of the halo mass on star formation history is {\em primary} and the influence
of stellar mass is secondary.
\end {enumerate} 
As we will discuss in the final section, our results indicate that the physical model for
determining the star formation histories of galaxies in the SAMs must be wrong.

In Fig.  \ref{fig:dssfr1},  we show the relation  between the quenched
satellite fraction  and the  projected distance to  the center  of the
scaled by $R_{\rm 200}$. Results are  shown in 3 halo mass bins.  Once
again we  see that the new  models produce results that  are much less
sensitive to resolution than than Fu13 models. As can be expected from
the results in the previous two figure, there are clear offsets between
the quenched  fractions in the models  and the data. The  slope of the
decrease in  quenched fraction as  a function of scaled  radius agrees
quite  well with  the observations  in high  mass haloes.  In low  mass
haloes, the  quenched fraction  rise too steeply  near the  center. The
steep rise towards the center is  stronger for the total specific star
formation rates compared  to the fibre specific  star formation rates,
and  it is  also stronger  for the  model that  includes ram-pressure
stripping of the  cold gas.  This suggests that  the central
density of hot gas in  lower mass  haloes is too  high or our rps model is too simple. 
Future constraints can be obtained from hydrodynamical simulations and the observed HI map of galaxies.

\begin{figure*}
\begin{center}
\includegraphics[scale=0.8]{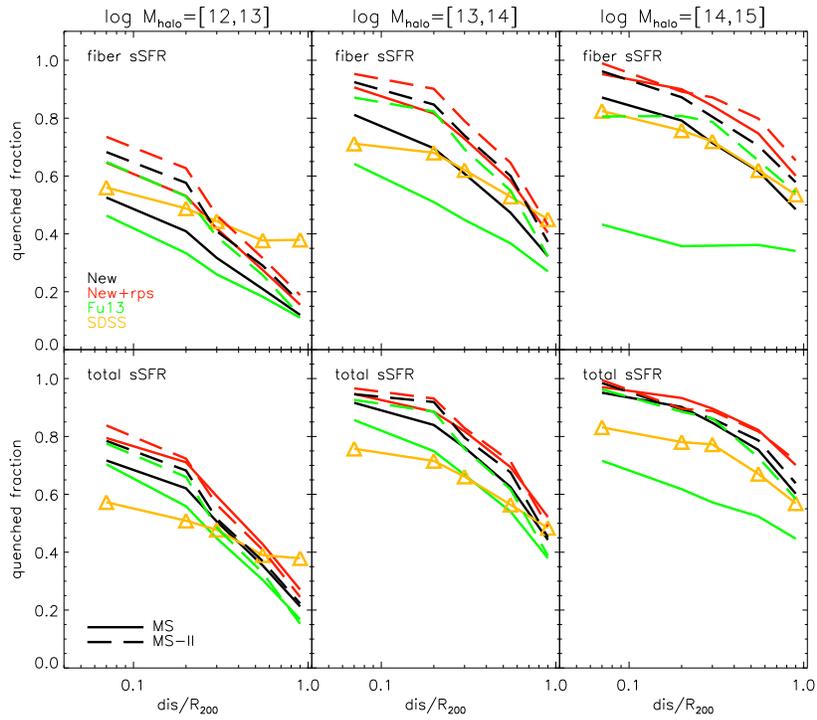}
\end{center}
\caption{The quenched satellite fraction versus the  projected
distance to the centre of the halo scaled by the virial radius of the halo. 
Results are shown in  3 different halo mass bins.}
\label{fig:dssfr1}
\end{figure*}
In  summary,  we  conclude  that  our  models  greatly  alleviate  the
resolution  problems  seen  in  the  Fu13 models,  but  are  still  in
significant discrepancies with the observational data remain.

\subsection{Comparison of the dependence of HI fraction and sSFR on environmental density}\label{chap:ssfr}

\begin{figure*}
\centering
\includegraphics[scale=1.0]{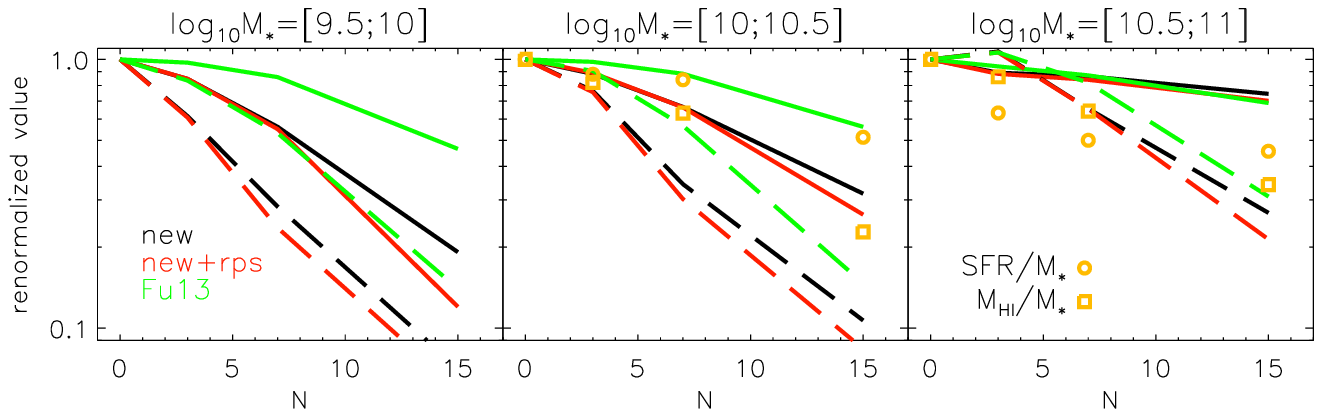}
\caption{The  relations between  the  values of  $M_{\rm HI}/M_*$ and  sSFR
  renormalized  to 1  at  N=0 as  a  function  of N,  where  N is  the
  environmental parameter  (see text for details).   Results are shown
  in two different  stellar mass bins. The right two  panels also show
  data from Fabello et al. 2012. The lines are model predictions, with
  solid  lines indicating  results for  fibre sSFR,  and dashed  lines
  indicating results for HI mass fraction .}
\label{fig:nssfr}
\end{figure*}

In this section, we compare model predictions of the HI mass fractions
of galaxies as a function  of environmental density to observations in
order  to  test  our  models of  ram-pressure  stripping.   Using  the
observational data from a few combined surveys (ALFALFA, GALEX, SDSS),
Fabello  et al.  (2012, hereafter  Fabello12) found  that both  the HI
fractions  and  sSFRs of  low-mass  galaxies  decline with  increasing
environmental density, but the HI fraction exhibits a steeper decline.
They compared the data with the  results of the Guo11 model, and found
that this model predicts  the opposite effect  -- the  sSFRs decline
more steeply with density than the HI mass fractions.

Following Fabello12, we  select galaxies  with $9.5<\log
(M_*/\rm{M_\odot})<11$,  and   compute  the   environmental  density
parameter  N,  defined   as  the  number  of   neighbours  with  $\log
(M_*/\rm{M_\odot})\geq9.5$ located  inside a cylinder of  $1 \rm{Mpc}$
in projected radius and with velocity difference less than $500 km/s$.
The    comparison   between    models   and    data   is    shown   in
Fig. \ref{fig:nssfr}. All curves have been normalized to 1 at N=0, and
the {\em relative} decrease in $\rm SFR/M_*$ and $\rm M_{HI}/M_*$ is plotted on
the y-axis. The  mass bins have been chosen to  match the stellar mass
bins shown in Fabello12.

Unlike the Guo11 model, the the HI mass fraction declines with density
more rapidly  than the specific star  formation rate for the  Fu13 and
for  the new  models. The  reason  for this  is because  in the  Guo11
models,  star formation  ceases  in the  whole galaxy  when the cold gas 
mass is lower than a threshold value.  This has the  consequence  that  
significant  gas  always  remains  in  passive
galaxies.  In  the Fu13  and our  new models, the  local star
formation  rate  is   determined  purely  by  the   local  gas  surface
density. As can be seen by comparing the red and the black curves, the
inclusion of ram-pressure stripping processes enhances the the decline
in HI mass fraction at large N, by only a small factor, indicating that
the star formation prescription rather than the treatment of stripping
is more important in understanding the observational results.

A more comprehensive test  of our ram-pressure stripping prescriptions
requires HI observations  of galaxies near the cores  of galaxy groups
and clusters.

\section{Conclusion and discussion}

The main results and findings of this paper can be summarized as follows:

\noindent (I) We include an analytic method to trace the properties of
the subhaloes of  satellite galaxies in haloes that  is independent of
the  resolution of  the  simulation.   This allows  us  to write  down
equations describing the  physical processes such as  SN feedback, gas
cooling,  gas  reincorporation  and   tidal  stripping  that  are  not
sensitive to whether  the galaxy is classified  as a Type 1  or Type 2
galaxy in the  simulation. The predicted gas  mass functions, quenched
satellite fractions and galaxy two-point correlation functions evaluated for the
two simulations  used in this  work, MS  and MS-II, agree  better than
in previous models.

\noindent  (II) We  include a  new  prescription to  describe the  ram
pressure stripping of cold gas in the galaxy.

We then  compare a variety of  results  with recent
observations. Our main conclusions are the following:

\noindent (i) Our new models allow for continued gas accretion and
SN feedback in all satellite galaxies. We improved the $\h2$ 
convergce for satellites in massive haloes which leads to more convergent quenched fraction of satellites in mass 
haloes. This has the effect of  
decreasing the clustering amplitude of low mass
galaxies on small scales, resulting in a  slightly better agreement with the observed
2pcf compared to previous models. 

\noindent (ii) We  show that ram pressure stripping  is most efficient
at  removing   the  cold   gas  from   low  mass   satellite  galaxies
($<10^{10.5}\rm{M_\odot}$)          in         massive          haloes
($>10^{13}\rm{M_\odot}$).  More than  60\%  of the  galaxies in  these
massive haloes have experienced significant ram-pressure stripping.

\noindent (iii) We  study the quenched fraction  of satellite galaxies
as a  function of stellar  mass at fixed halo  mass, and as a  function of
halo mass  at fixed stellar  mass.  We find  significant discrepancies
between our model  predictions and observations.  At the fixed halo mass,
the quenched fraction of satellites does not depend on stellar mass in
the models. This  is in contradiction with observations where the
quenched fraction always increases with  stellar mass.  The net effect
of  this discrepancy  is that  there are  too many  low mass  quenched
satellites and  too few  high mass quenched  satellites in  the models
compared to the data.

\noindent  (iv) We  study the  decrease  in the  quenched fraction  of
satellite galaxies  as function of  of projected radial  distance from
the center  of the halo.  The slope of  the decrease agrees  well with
observations in high mass haloes, but  is much steeper than observed in
low mass  haloes.  This problem  is worse  for models that  include the
ram-pressure stripping of cold gas,  indicating that the predicted hot
gas densities in the centers of lower mass haloes may be too high.

\noindent (iv)  Our new  models are able  to reproduce  the relatively
stronger  decrease in  HI gas  mass fraction  as a  function of  local
environmental density  compared to specific star  formation rate first
pointed out by Fabello12.

Our study  in this paper shows  that the long-standing problem  of the
over-prediction of red  satellite galaxies is still not  solved in the
current version of the L-Galaxies model.  Recently Sales et al. (2015)
studied the color of satellite  galaxies in the Illustris cosmological
simulations.   They found  that  their simulation  produces more  blue
satellites and  proposed that  the main reason  SAMs fail  to reproduce
satellite colors  is that there is  too little cold gas  in satellites
before  they  are accreted.   We  regard  this scenario  as  unlikely,
because  as we  have shown,  our SAMS  produce gas  mass functions  in
excellent  agreement  with  observations  at  the  present  day.  Fu13
demonstrated that  the $\h2$  mass functions  of galaxies  evolve very
strongly with redshift.  Although keeping  more cold gas in satellites
before accretion could in principle solve  the problem of too many red
satellites, it  would likely violate other  observational constraints.
Kang (2014) have also shown that  the stellar mass growth in satellite
galaxies after infall should not be significant, otherwise there would
be too many low-mass galaxies  in massive clusters, and the clustering
amplitude would  be too  high clustering on  small scales.   Wetzel et
al.  (2013) also  found  from SDSS  that the  stellar  mass growth  of
satellite galaxies in constrained to be less than 60\% on average.

Recently, it has been shown that a large fraction of low mass galaxies do
not experience continuous star formation histories, but have undergone
a significant bursts of star  formation (Kauffmann 2014). What exactly
causes these bursts  is not yet understood, but in  all likelihood all
these  results indicate  that gas  accretion processes  onto low  mass
field  galaxies are  more complex  than assumed  in the current semi-analytic
models.   We note  that  the net  effect of  a  bursty star  formation
history is to  produce a sSFR distribution with a  tail that is skewed
towards  {\em low  specific star  formation rates},  because the  duty
cycle of the burst phase when the galaxy is forming stars very rapidly
is short.  An extra tail  of low sSFR  galaxies at low  stellar masses
would bring  the model central  galaxy quenched fractions  into better
agreement with observations, as seen  in the rightmost panel of Figure
8.   It would  also  lower  the specific  star  formation  rates of  a
significant fraction of galaxies being accreted as satellites, perhaps
causing them  to exhaust  their existing  gas reservoirs  more slowly.
Another issue that  was pointed out in a paper  by Kauffmann (2015) is
that  there  is a  correlation  between  low  mass galaxies  that  are
quenched and  the quenched fraction  of their neighbours  that extends
over  very  large scales  and  that  is  not currently  reproduced  by
semi-analytic models or by the Illustris simulation.

Until  we  have a  model  that  reproduces  basic trends  in  quenched
fractions as a function of both stellar  mass and halo mass, it will be
difficult to come to robust conclusions as to whether our treatment of
ram-pressure stripping of cold gas is a significant improvement to the
models.   We  have presented  some  tentative  evidence that  hot  gas
densities at the centers of lower mass haloes ($\sim 10^{12} M_{\odot}$)
are  smaller  than  predicted  by our  models,  because  the  quenched
fractions  at  the  very  centers  of these  haloes  are  too  high  in
comparison with  observations. One way  to reduce the central  hot gas
densities  is through  radio AGN  feedback processes.  In the  current
implementation  of  radio AGN  feedback,  the  cooling rate  onto  the
central  galaxy   is  reduced,   but  the  gas   distribution  remains
unaffected,  which is  not  physically reasonable.  At  this level  of
detail, however,  hydrodynamical simulations may provide  a better way
forward.

\section*{Acknowledgements}

We thank the anonymous referee for useful comments. We also thank Bruno Henriques, Yannick Bah\`e, Qi Guo and Simon White for helpful discussions.
This  work  is  supported  by   the  973  program  (No.  2015CB857003,
2013CB834900), NSF of Jiangsu (No.BK20140050), the NSFC (No. 11333008,
111303072,U1531123) and the Strategic Priority Research Program the emergence of
cosmological structure  of the CAS (XDB09000000).  YL acknowledges the
support   and   hospitality   by    the   Max-Planck   Institute   for
Astrophysics. JF acknowledges the support by the Opening Project of Key 
Laboratory of Computational Astrophysics, National Astronomical Observatories, CAS.

\def\apj{ApJ}
\def\apjl{ApJL}
\def\apjs{ApJS}
\def\aj{AJ}
\def\aap{A\&A}
\def\araa{ARA\&A}
\def\aapss{A\&AS}
\def\mnras{MNRAS}
\def\nature{Nature}
\def\apss{Ap\&SS}
\def\pasp{PASP}

{}

\end{document}